\documentstyle[12pt,epsf]{article}
\begin{document}
\begin{flushright} {OITS 688}\\
March 2000
\end{flushright}
\vspace*{1cm}

\begin{center}  {\large {\bf Fluctuation of Voids in Hadronization at
Phase Transition}}
\vskip .75cm
 {\bf  Rudolph C. Hwa and  Qing-hui Zhang}

 {\bf Institute of Theoretical Science and Department of
Physics\\ University of Oregon, Eugene, OR 97403-5203\\}
\end{center}

\begin{abstract}
Starting from the recognition that hadrons are not produced smoothly at
phase transition, the fluctuation of spatial patterns is investigated by
finding a measure of the voids that exhibits scaling behavior.  The Ising
model is used to simulate a cross-over in quark-hadron phase transition.
A threshold in hadron density is used to define a void.  The dependence of
the scaling exponents on that threshold is found to provide useful
information on some properties of the hadronization process.  The
complication in heavy-ion collision introduces the possibility of
configuration mixing, which can also be studied in this approach.
Numerical criteria on the scaling exponents have been found that can be
used to discriminate phase-transition processes from other hadronization
processes having nothing to do with critical phenomena.

\end{abstract}

\section{Introduction}
One of the generic properties of second-order phase transition is
clustering.  In the case of a magnetic system, it means that there are
regions of all sizes, in which spins point in the same direction, and
that the probability of having a cluster of a certain size satisfies a
scaling law.  The implication of those properties for quark-hadron phase
transition (PT) is that hadronization does not occur uniformly at the
critical temperature $(T_c)$.  At any instant during the entire course
of the hadronization process of a quark-gluon plasma system, there are
then clusters of hadrons separated by regions of no hadrons.  In the
case of heavy-ion collisions it is the surface of the expanding cylinder
that is around $T_c$, and the clusters are to appear on the
two-dimensional cylindrical surface.  We call the non-hadronic regions
between the clusters voids.  To detect voids will be a sign of the
second-order PT.  In this paper we study the properties of the voids and
discuss how to identify their properties in heavy-ion experiments.

It should be understood that voids are not frozen on a plasma surface for
all times.  If one could take instantaneous pictures of the cylindrical
surface at time intervals of 1 fm/c apart, one would see the
clustering patterns to fluctuate from picture to picture.  Simulation on
a 2D lattice using the Ising model shows that at $T_c$ the clustering
patterns differ from one configuration to another.  An example of the
clusters of hadrons formed can be seen in Fig.\ 2 of Ref.\ \cite{1}.
There are regions of high hadron density, low hadron density, and no
hadrons.  A region without hadrons (i.e. a void) consists of quarks and
gluons in the deconfined state at a particular instant in time; they are
likely to form hadrons a little later in the evolution process.  The
reason for the voids to exist is that at the critical point the system
is torn between being in the ordered state of confinement and the
disordered state of deconfinement.  It is this tension of coexistence of
the two states at $T_c$ that is responsible for the many interesting
behaviors of critical phenomena
\cite{2}.  Thus if the plasma volume created in a heavy-ion collision is
hot in the interior and cools to $T_c$ on the surface, a second-order PT,
which is assumed here, would imply that the hadrons are not produced
smoothly in time or space.  Whereas hadronic clusters in the
$\eta$-$\phi$ space may be hard to quantify, voids are relatively easy
to define, though not trivially.  Our first problem will be to
characterize the voids.  To identify them in the theoretical laboratory
is, however, much simpler than to find them in the experimental data.
The latter problem will also be addressed in this paper.

A simple way to appreciate voids is to examine the exclusive distribution
in rapidity space for a hadronic collision process.  Since the rapidity of
each produced particle can be precisely determined, one can calculate the
rapidity gaps between each pair of neighboring particles.  Those rapidity
gaps are the 1D version of the voids in 2D.  An extensive
consideration of the properties of gaps in hadronic processes is given
in Ref.\ \cite{3}.  Although it is easy to gain a mental picture of gaps
in hadronic collisions, it is nontrivial to define a measure of voids in
heavy-ion collisions, in which the hadronization process extends over a
long period of time.

To gain some physical insight into the fluctuation phenomenon in particle
production at the critical point, it is helpful to consider the production
of photons at the threshold of lasing.  The physics of single-mode lasers
being completely understood, their operation near the threshold is known to
behave as in a second-order PT \cite{4}.  When the pump parameter is set at
the threshold of lasing, the system does not continuously produce photons
as a function of time.  The photons are produced in spurts with gaps of
quiescence between spurts.  Such fluctuations have been measured and the
result in terms of factorial moments \cite{5} agree with the prediction
bases on the Ginzburg-Landau description of PT \cite{6}.  Those gaps in the
time series in the photo-count problem are similar to the voids in the
hadron-count problem in heavy-ion collisions.  Since lattice QCD cannot be
applied efficiently to the simulation of spatial patterns in the PT
problems, we make use of the universal features of the Ginzburg- Landau
theory and, in particular, employ the 2D Ising model to simulate the
hadronic clusters in the  $\eta$-$\phi$ plane.  This approach to the
problem was initiated in \cite{7}, where the scaling properties of cluster
production were examined.  More recently, an effort was made to find
observable critical behavior in quark-hadron PT \cite{1}.  Here, we use
the same formalism to study the properties of voids and search for
observable measures that can signal PT in heavy-ion collisions.

\section{Hadron Production in the Ising Model}

It is known through studies in lattice QCD that the nature of the phase
transition depends on the number of flavors and the quark masses \cite
{8}.  For $m_u=m_d=0$, the PT is first-order for low $m_s$, but
second-order at high $m_s$.  For nonzero $m_u$ and $m_d$, the former region
remains first-order (but for even smaller values of $m_s$), while the
latter region of high $m_s$ becomes a cross-over.  The two regions are
separated by a phase boundary that is second-order and has possibly the
Ising critical exponents \cite{9}.  For realistic quark masses we are
probably in the region of the cross-over, which is what we shall assume in
this study.

A cross-over means that no calculable or measurable quantities and their
derivatives undergo any discontinuity as $T$ is varied across $T_c$.  In
the 2D Ising model the situation is like having a small external
magnetic field so that the average magnetization of the system varies
smoothly across $T_c$.  The phase diagrams of the order parameter versus
$T$ are very similar for the Ising model and the QCD problem.  Since
lattice QCD is so much more difficult to study compared to the Ising
model, we shall hereafter concentrate only on the use of the Ising model
to simulate hadron production.

Following the formalism already described in Refs. \cite{1,7} let us
briefly summarize how hadron density is defined on the Ising lattice in
2D.  For a lattice of size $L \times L$ with each site having spin
$\sigma_j$, we define the spin aligned along the overall magnetization
$m_L = \sum_{j \in L^2} \sigma_j$ by
\begin{eqnarray}
s_j = sgn (m_L) \sigma_j \quad
\label{1}
\end{eqnarray}
where $sgn(m_L)$ stands for the sign of $m_L$.  We then define the spin
of a cell of size $\epsilon \times \epsilon$ at location $i$ by
\begin{eqnarray}
 c_i = \sum_ {j \epsilon A_i} s_j \quad
\label{2}
\end{eqnarray}
where $A_i$ is the cell block of $\epsilon^2$ sites at $i$.  Since $c_i$
averages over all site spins in a cell, it should approach zero at high
$T$ for which the system is in a disordered state, and should approach
$\epsilon^2$ as $T
\rightarrow 0$ for which the system is in an ordered state.  Note that even
in the absence of an external magnetic field, $c_i$ approaches
+$\epsilon^2$ at low $T$, never $-\epsilon^2$, because of our definition
of
$s_j$.  Thus unlike the average magnetization $\left<m_L\right>$ of the
usual Ising model without external field, which is zero for all $T>
T_c$, the average cell spin $\left<c_i\right>$ varies smoothly from high
to low
$T$, similar to the behavior of $\left<m_L\right>$ in the presence of
external field.

The hadron density, being proportioned to the order parameter, can now be
defined as
\begin{eqnarray}
 \rho_i = \lambda\, c_i^2 \,\theta (c_i) \quad,
\label{3}
\end{eqnarray}
where $\lambda$ is an unspecified factor relating the lattice spins to
the number of particles in a cell.  In any configuration $c_i$ may still
fluctuate from cell to cell.  We identify only the positive $c_i$ cells
with hadron formation, and associate the hadron density with $c_i^2$,
just as the order parameter in the Ginzburg-Landau formalism is
associated with the square of the Ising spins \cite{2}.  If
$\left<\rho\right>$ denotes the average density, i.e., $\rho_i$ averaged
over all cells on the lattice and over all configurations, then the
dependence of $\left<\rho\right>$  on $T$ is as shown in Fig.\ 1.  It is
typical of a cross-over, shown in Fig.\ 1 of Ref.\ \cite{9} for small
quark masses.  The essence of that dependence is that
$\left<\rho\right>$ decreases precipitously but smoothly, as
$T$ is increased across $T_c$, but remains nonzero for a range of $T$ above
$T_c$.  Evidently, we have succeeded in simulating a cross-over without
the explicit introduction of an external field in the Ising Hamiltonian
\begin{eqnarray}
H = -J \sum_{\left<ij\right>} \sigma_i \sigma_j \quad.
\label{4}
\end{eqnarray}

Since the dependence of $\left<\rho\right>$ on $T$ is smooth, it is a
nontrivial problem to determine the precise value of $T_c$.  That has
been done in Ref. \cite{7} by examining the scaling behavior of the
normalized factorial moments $F_q$.  It is found that $F_q$ behaves as
$M^{\varphi_q}$ only at $T = 2.315$ (in units of $J/k_B$), where
$M$ is the number of bins on the lattice.  Since the critical point is
characterized by the formation of clusters of all sizes in a scale
independent way, we identify the critical temperature at $T_c = 2.315$.
Indeed, it has been shown in Ref. \cite{1} that in the neighborhood of
$T_c$ with $T\leq T_c$, $\left<\rho\right>$ behaves as
\begin{eqnarray}
\left<\rho\right> - \left<\rho_c\right>  \propto
\left(T_c-T\right)^\eta,
\qquad
\eta=1.67 \quad,
\label{5}
\end{eqnarray}
where $\left<\rho_c\right>$ is $\left<\rho\right>$ at $T_c$.  There exist
other measures that exhibit critical behaviors,
 $\left(T_c-T\right)^{-\zeta}$, with
negative exponents.  They are discussed in \cite{1}.

\section{Scaling Behavior of Voids}

We choose to work with a lattice having $L = 256$ and cells having
$\epsilon = 4$.  Thus the total number of cells on the lattice is $N_c =
(L/\epsilon)^2 = 64^2$.  We divide the lattice into bins of size
$\delta^2$ so that each bin can contain $\nu = (\delta/\epsilon)^2$
cells and the lattice can contain $M =(L/\delta)^2$ bins.  The average
density of hadrons in a bin is therefore
\begin{eqnarray}
\bar{\rho}_b = {1 \over \nu} \sum^\nu_{i=1} \rho_i \quad ,
\label{6}
\end{eqnarray}
where $b$ denotes the $b\rm th$ bin.  Near $T_c$, $\bar{\rho}_b$
fluctuates from bin to bin, especially for small $\delta$.  We define a
bin to be ``empty'' when
\begin{eqnarray}
\bar{\rho}_b < \rho_0 \quad ,
\label{7}
\end{eqnarray}
where $\rho_0$ is a floor level greater than zero.  This criterion is
chosen to eliminate the effect of small fluctuations on gross
behavior.  That is, for the purpose of defining a void, hadron clusters
are counted only when the hadron density is above a threshold
$\rho_0$.  Bins with very low hadron density, i.e., where (\ref{7})
holds, are then regarded as empty.  A void is a contiguous collection
of empty bins.  Fig.\ 2 illustrates a pattern of voids in a
configuration generated at $T_c$ for $M = 24^2$ and $\rho_0 = 20$ (in
units of $\lambda$).  An open square indicates an empty bin, while a
black square contains hadrons with $\bar{\rho}_b \geq \rho_0$.  In
that configuration there are 26 voids, the sizes of which are 76, 35,
9, 8, 7, $\cdots$ in descending order.  It should be recognized that
the maximum hadron density that a cell can have is $\rho_{\rm max} =
\left(\epsilon ^2\right)^2 = 256$, so $\rho_0 = 20$ represents a
threshold that is less than 8\% of the maximum.

Let $V_k$ be the size of the $k\rm th$ void (in units of bins).  That is,
let
\begin{eqnarray}
V_k = \sum_{\left<b\right>_k} \theta \left(\rho_0 -\bar{\rho}_b\right)
\quad ,
\label{8}
\end {eqnarray}
where $\left<b\right>$ implies a sum over all empty
bins that are connected to one another by at least one side; $k$ simply
labels a particular void.  We can then define $x_k$ to be the fraction
of bins on the lattice that the $k\rm th$ void occupies:
\begin{eqnarray}
x_k = V_k / M.
\label{9}
\end {eqnarray}
For each configuration we thus have a set ${\cal S} = \{x_1, x_2,
\cdots\}$ of void fractions that characterizes the spatial pattern.

Since the pattern fluctuates from configuration to configuration, ${\cal
S}$ cannot be used to compare patterns in an efficient way.
For a good measure to facilitate the comparison, let us first define the
moments $g_q$ for each configuration
\begin{eqnarray}
g_q = { 1 \over m } \sum^m_{k=1} x_k^q \quad ,
\label{10}
\end{eqnarray}
where the sum is over all voids in the configuration, and $m$ denotes the
total number of voids.  We then define the normalized $G$ moments
\begin{eqnarray}
G_q = g_q / g^q_1 \quad ,
\label{11}
\end{eqnarray}
which depends not only on the order $q$, but also on the total number of
bins $M$.  Thus by definition $G_0 = G_1 = 1$.  This $G_q$ is defined in
the same spirit as that in \cite{3} for rapidity gaps, but they are not
identical because the $x_k$ here for voids do not satisfy any sum rule.  It
is also unrelated to the $G$ moments defined earlier \cite{10} for fractal
analysis.  Now, $G_q$ as defined in Eq. (\ref{11}) is a number for every
configuration for chosen values of $q$ and $M$.  With $q$ and $M$ fixed,
$G_q$ fluctuates from configuration to configuration and is our
quantitative measure of the void patterns, which in turn are the
characteristic features of phase transition.

In Fig.\ 3 we show the probability distribution of $G_q$ for $q = 6$ and $M
= 36^2$ and $\rho_0 = 20$ at three different values of $T$.  The Wolff
algorithm has  been used in the Monte Carlo simulation to reduce the
correlation between configuration \cite{2, 11}.  The distribution in
Fig.\ 3 and other quantities to be calculated below are the results
obtained using
$5\times 10^3$ uncorrelated configurations.  Since the value of $G_q$
fluctuates widely from configuration to configuration covering a range
that exceeds 3 orders of magnitude, we have plotted the distribution in
$lnG_q$.  Since both the mean and the dispersion of
$lnG_q$ shown in Fig.\ 3 vary significantly with $T$, and with $q$ and $M$
not shown in Fig.\ 3, it is necessary for us to search for simple
regularities in the nature of the fluctuations of $G_q$.

Our first step in that search is to study the $M$ dependence of the average
of $G_q$ over all configurations, i.e.,
\begin{eqnarray}
\left<G_q\right> = { 1 \over \cal N } \sum^{\cal N}_{e=1} G_q^{(e)} \quad
,
\label {12}
\end{eqnarray}
where the superscript $(e)$ denotes the $e\rm th$ event
(or configuration) and
$\cal N$ is the total number of events.  In Fig.\ 4 we show
$\left<G_q\right>$ versus $M$ in a log-log plot for $T=T_c$, $2\leq q
\leq 8$, and $\rho_0 = 20$.  We find very good linear behavior;
consequently, we may write
\begin{eqnarray}
\left<G_q\right> \propto M ^{\gamma_q} \quad  .
\label{13}
\end{eqnarray}
This scaling behavior implies that voids of all sizes occur at PT.  Since
the moments at different $q$ are highly correlated, we expect the scaling
exponent $\gamma_q$ to depend on $q$ in some simple way.  Fig.\ 5 shows
the dependence of $\gamma_q$ on $q$, and we find remarkable
linearity.  Thus we may write
\begin{eqnarray}
\gamma_q = c_0 + c\,q \quad ,
\label{14}
\end{eqnarray}
where $c = 0.8$.  There is no obvious reason why the $q$-dependence of
$\gamma_q$ should be so simple.  We should regard (\ref{14}) only as a
convenient parameterization of $\gamma_q$ that allows us to focus on $c$
as a numerical description of the scaling behavior of the voids at PT.

It is evident from Fig.\ 3 that studying the behavior of $\left<G_q\right>$
extracts only a limited amount of information about the distribution
$P(G_q)$.  The fluctuation of $G_q$ from event to event can be quantified
by the various moments of $G_q$
\begin{eqnarray}
C_{p,q} = {1 \over {\cal N}} \sum^{\cal N}_{e=1} \left(G^{(e)}
_q\right)^p =
\int dG_q \, G_q^p \, P(G_q) \quad ,
\label {15}
\end{eqnarray}
among which $\left<G_q\right>$ corresponds only to $C_{1,q}$.  Instead of
examining a collection of $C_{p,q}$ for various values of $p$, we consider
the derivative of $C_{p,q}$ at $p=1$ \cite {3,12}, and define
\begin{eqnarray}
S_q =  {d \over dp} C_{p,q}\big|_{_{p=1} }=  \left<G_q ln G_q\right>
\quad ,
\label {16}
\end{eqnarray}
where $\left<\cdots\right>$ stands for averaging over all
configurations.  Despite its appearance, $S_q$ is not entropy, but is a
measure of the fluctuations of $G_q$, when compared with
$\left<G_q\right> ln
\left<G_q\right>$.

In Fig.\ 6 we show the power-law behavior of $S_q$ at $T_c$
\begin{eqnarray}
S_q \propto M^{\sigma_q} \quad ,
\label{17}
\end {eqnarray}
where the scaling exponents  $\sigma_q$ are the slopes of the straight
lines in the figure.  Because of Eq.(\ref {13}), $\left<G_q\right>
ln
\left<G_q\right>$ is not power behaved, so $S_q - \left<G_q\right> ln
\left<G_q\right>$ would not have a scaling behavior as in
Eq.(\ref{17}).  For that reason we focus on the simple properties of
$S_q$.  The dependence of $\sigma_q$ on $q$ is shown in Fig.\ 7, where a
remarkable linear behavior is found.  We use the parameterization
\begin{eqnarray}
\sigma_q = s_0 + s\,q
\label {18}
\end{eqnarray}
and find $s=0.76$.

Among the quantities that are under our control in the analysis, we have
studied the dependences on $M$ and $q$.  The remaining such quantity is
$\rho_0$, while $T$ is beyond experimental control, although it can be
varied in the simulation.  We now study the dependence of $c$ and $s$ on
$\rho_0$ and $T$, which are shown in Figs.\ 8 and 9.  Evidently, at
higher values of $\rho_0$ the dependence on $T$ are more pronounced than
at $\rho_0 = 20$.  Similar behavior has been found for $c_0$ and $s_0$,
defined in Eqs. (\ref{14}) and (\ref{18}).  This result is very interesting,
and provides a possible avenue toward learning more about the nature of
PT in a realistic heavy-ion experiment.

If a quark-gluon plasma is formed in a heavy-ion collision, the expanding
cylinder has high $T$ in the interior and low $T$ on the surface.  Our
modeling has been to investigate the properties of hadronization on the
surface.  Since the PT is a smooth cross-over in the neighborhood of
$T_c$, and also since hydrodynamical flow can lead to local fluctuations in
temperature and radial velocity on the surface, it is realistic to expect the
hadrons to form in a small range of $T$ around $T_c$, a possibility that
cannot be controlled experimentally nor excluded theoretically.  To learn
whether the hadronization takes place over a range of $T$, we suggest on
the basis of Figs. 8 and 9 to use $\rho_0$ as a device to probe the
properties of the PT.

In an analysis of the experimental data one can use a phenomenological
density threshold to play the role of $\rho_0$ and vary it in the
determination of the patterns of voids.  From our study we have learned
that for a wide range of $\rho_0$ the values of $c$ and $s$ are not
independent of $T$.  It means that if the PT occurs over a range of $T$
so that the hadrons detected are formed at various $T$ around $T_c$
even within one event, then for $\rho_0$ in that range of nonuniform $c$
and $s$ the dependences of $\left<G_q\right>$ and $S_q$ on $M$ would
not exhibit simple power-law behavior as in Figs.\ 4 and 6, since
nonuniform values of $\gamma_q$ and $\sigma_q$ for any $q$ would
lead to $M$-dependences that are not simple power-laws. If, by
varying $\rho_0$ to a point that corresponds to 20 in this study, simple
scaling behaviors can be found for $\left<G_q\right>$ and $S_q$ with $c$
and $s$ having roughly the values 0.8 and 0.76 respectively, then we can
be assured that the PT is a cross-over of the type that we have studied
here in the Ising model.  The signature for having a unique $T_c$ for
hadronization is that the scaling behavior persists at any $\rho_0$ and
the lowest values of $c$ and $s$ are at 0.7 and 0.67, respectively, when
$\rho_0 = 50$.

\section{Configuration Mixing}

The possibility of hadronization occurring at a narrow range of
temperatures discussed at the end of the previous section is not the only
complication that may occur in a heavy-ion collision process.  A
configuration that we simulate on the Ising lattice corresponds to the
cluster and void pattern of one instant (with uncertainty 1 fm/c) in the
hadronization history of a plasma cylinder.  The final state of a
collision process registered at the detector is a collection of all the
particles produced throughout the whole evolution process in excess of
10 fm/c.  The clusters and voids produced at different times overlap one
another in the integration process, resulting in a smooth spatial
distribution.  Thus to identify the clustering patterns in an
experiment, it is necessary to make cuts in the phase space.  Since
$\eta$ and $\phi$ are needed to to exhibit the 2D patterns, $p_T$ is the
only remaining variable, in which cuts can be made.  Since there is some
correlation between $p_T$ and the evolution time, the selection of
particles having their $p_T$ lying in a very narrow interval, $\Delta
p_T$, has the effect of selecting a small interval, $\Delta \tau$, in
evolution time
\cite{1}.  However, the correspondence is not one-to-one.  In every
$\Delta p_T$ interval, patches of hadrons produced at neighboring times
can contribute.  That is what we mean by configuration mixing:  the
experimental configuration detected in a small $\Delta p_T$ interval may
be a mixture of parts of configurations produced at different times, the
latter being the pure configurations that we simulate on the Ising lattice.

To simulate a mixed configuration, we make the following choice for
definiteness.  We divide the lattice into four quadrants.  In each of the
quadrants we place the corresponding quadrant of a new and
independent configuration so that the mixed configuration consists of four
parts of four configurations.  On $5\times 10^3$ such mixed
configurations we then performed the same analysis as in the preceding
section.  The results are summarized in Figs.\ 10 and 11 where, for
comparison, the straight lines are reproduced from those in Figs.\ 5 and
7 for the pure configurations.  The squares are the results for
$\gamma_q$ and
$\sigma_q$ calculated from the mixed configurations.  Evidently,
configuration mixing does not introduce any discernible deviation from
the results of the pure configurations.  The agreement being so good, we
see no point in trying out other ways of mixing.  The implication of the
result is remarkable, but not surprising.  The cluster and void patterns
fluctuate so much that it does not matter whether some pieces of the
patterns come from different configurations, provided that the
appropriate measure of the fluctuations is extracted.  What we have
extracted is the scaling behavior in $M$.  The scaling exponents are
then found to be independent of the configuration mixing.

\section{Conclusion}

We have shown in this paper that the study of voids can be very fruitful in
finding
signals of quark-hadron phase transition in heavy -ion collisions.  The
use of 2D Ising model has been effective in simulating a cross-over in the
hadronization process.  $G_q$ moments have been defined to quantify the
dependence of the voids on the bin sizes.  The scaling behavior that has
been found provides an efficient way to use the scaling exponents
$\gamma_q$ and $\sigma _q$ to characterize the properties of the phase
transition.

The temperature at which hadronization occurs is not under experimental
control.  We have found a way to learn whether hadronization occurs at a
range of $T$ or at a unique $T$.  That is achieved by varying the density
threshold $\rho_0$. A bin whose average density is $<\rho_0$ is
identified as belonging to a void.  The quantity $\rho_0$ is under the
control of the analyst of the experimental data.  If by varying $\rho_0$
one finds that a scaling behavior can be tuned out, i.\ e.\, the
power-law dependence on
$M$ becomes invalid for a range of $\rho_0$, then hadronization does not
occur at a unique $T$.  On the other hand, if scaling remains manifest
for a range of $\rho_0$, then there is only one temperature at which
hadrons are formed.

Even in the case where hadronization takes place in a range of $T$, it is
possible to tune $\rho_0$ to a value where strict scaling can be
observed.  Then $\gamma_q$ and $\sigma _q$ provide the slope parameters
$c$ and $s$ that can be checked as numerical constants characteristic of
the cross-over PT.  Since there are no numerical inputs in our
analysis, the values $c = 0.8$ and $s = 0.76$ are predictions in this
study.  Experimental verification of those numbers would, of course, lend
significant support to this line of study.  If the experimental numbers
for $c$ and $s$ turn out not to have those values, or if there is no
scaling behavior at all, one could conclude that the hadrons are formed
without the system having gone through a phase transition.

We have further found that the study of the scaling behavior of the void
moments has the additional virtue of being independent of configuration
mixing.  That property strengthens the argument that a small
$\Delta{p_T}$ cut in the data can provide us with a window to look into
the hadronization process at a small $\Delta \tau$ time interval where
hadron clusters and voids are formed.  The effect of randomization by
the possible hadron gas in the final state on the scaling behavior has
already been found in \cite{1} to be minimal.  Thus, we have here a
promising procedure to investigate the properties of quark-hadron phase
transition that should be undertaken by the heavy-ion experiments.

\begin{center}
\subsubsection*{Acknowledgment}
\end{center}

We are grateful to  Z.\ Cao and Y.F.\ Wu for providing us with the
original code on the Ising model.  This work was supported, in part, by
U.\ S.\ Department of Energy under Grant No.\ DE-FG03-96ER40972.


\vspace{2cm}
\newpage
\section*{Figure Captions}

\begin{description}
\item[Fig.\ 1]Average hadron density (in units of $\lambda$) versus
temperature (in units of $J/k_B$)

\item[Fig.\ 2]Spatial pattern of a configuration on the 2D lattice at
$T_c$. Open squares indicate voids and filled squares indicate hadrons
with $\rho_0=20$.

\item[Fig.\ 3]Probability distributions of ln$G_q$ (with $q=6$) at two
different
$T$.

\item[Fig.\ 4]Scaling behavior of $\langle G_q\rangle$ vs $M$ at $T_c$.

\item[Fig.\ 5]The dependence of $\gamma_q$ on $q$.

\item[Fig.\ 6]Scaling behavior of $\langle S_q\rangle$ vs $M$ at $T_c$.

\item[Fig.\ 7]The dependence of $\sigma_q$ on $q$.

\item[Fig.\ 8]The dependences of the slope parameter $c$ on $\rho_0$ and
$T$.

\item[Fig.\ 9]The dependences of the slope parameter $s$ on $\rho_0$ and
$T$.

\item[Fig.\ 10]A comparison of $\gamma_q$ between the mixed
configurations (squares) and the pure configurations (straightline taken
from Fig.\ 5).

\item[Fig.\ 11]A comparison of $\sigma_q$ between the mixed
configurations (squares) and the pure configurations (straightline taken
from Fig.\ 7).

\end{description}

\begin{figure}[t]\epsfxsize=14cm \epsfysize=14cm
\centerline{\epsfbox{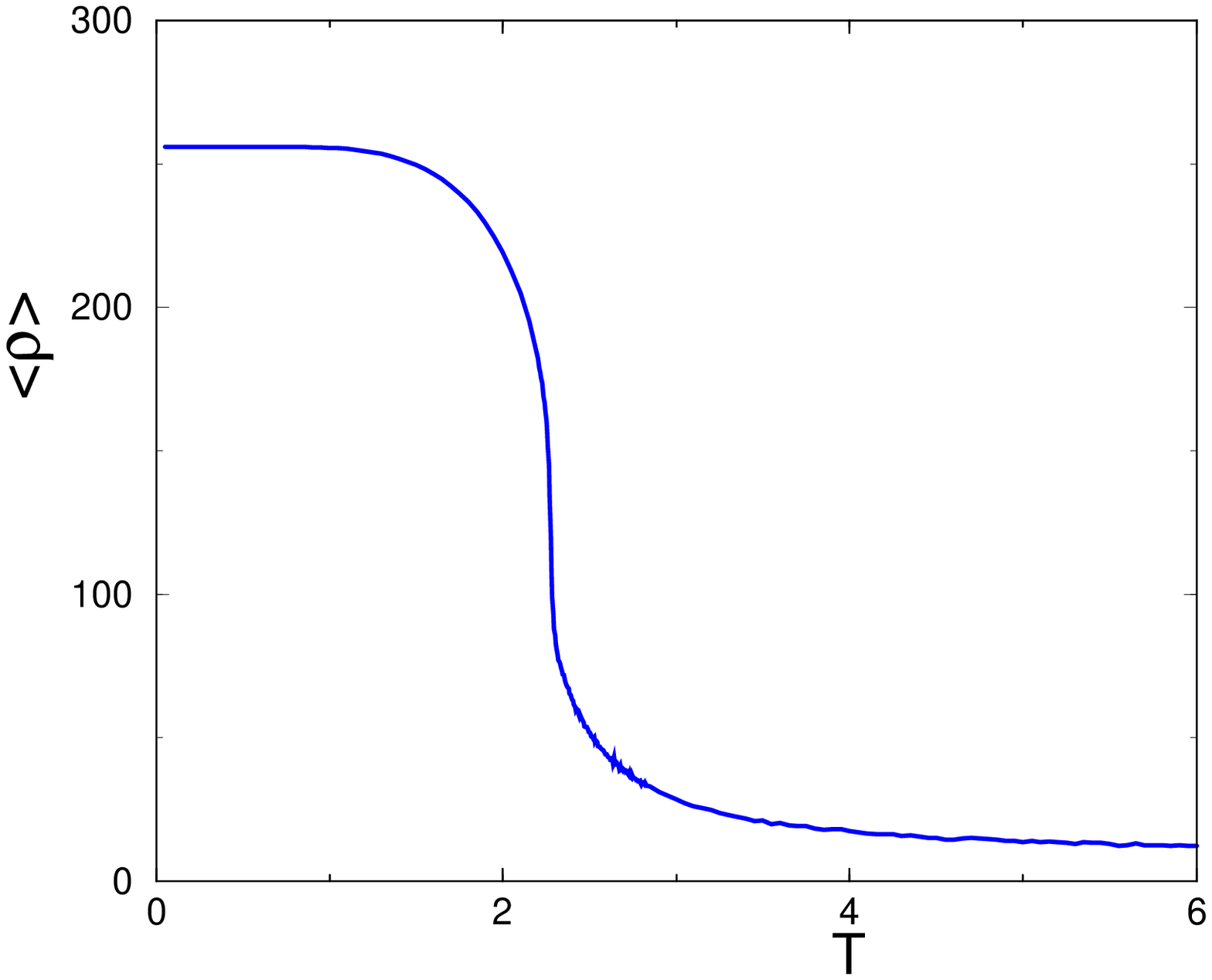}}
\vskip 1.5 cm
\centerline{\bf \large ~~~~~~~~~~Fig.1}
\vskip -0.5 cm
\end{figure}

\newpage
\begin{figure}[t]\epsfxsize=14cm \epsfysize=14cm
\centerline{\epsfbox{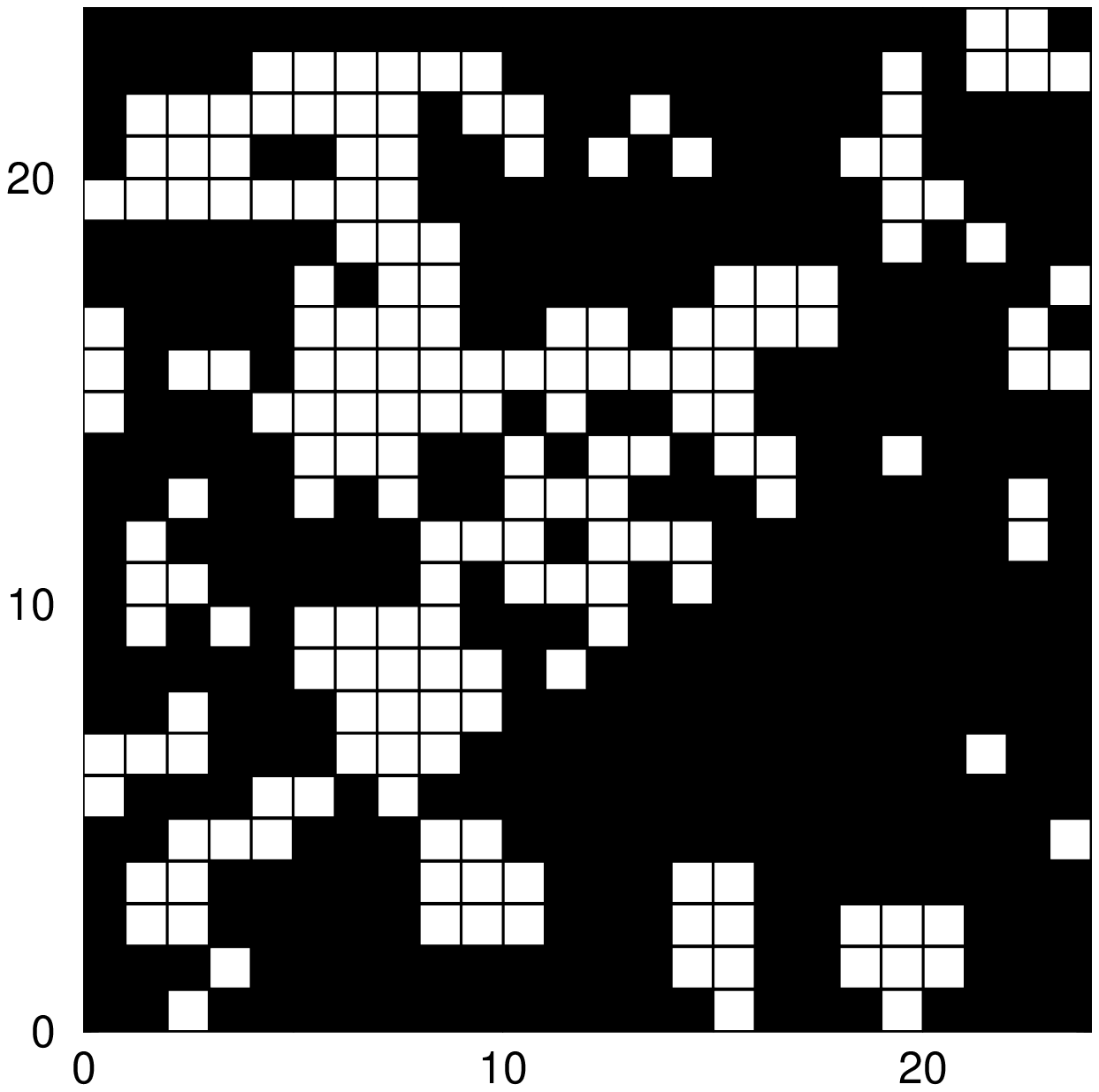}}
\vskip 1.5 cm
\centerline{\bf \large ~~~~~~~~~~~Fig.2}
\vskip -0.5cm
\end{figure}

\newpage
\begin{figure}[t]\epsfxsize=16.5cm \epsfysize=14cm
\centerline{\epsfbox{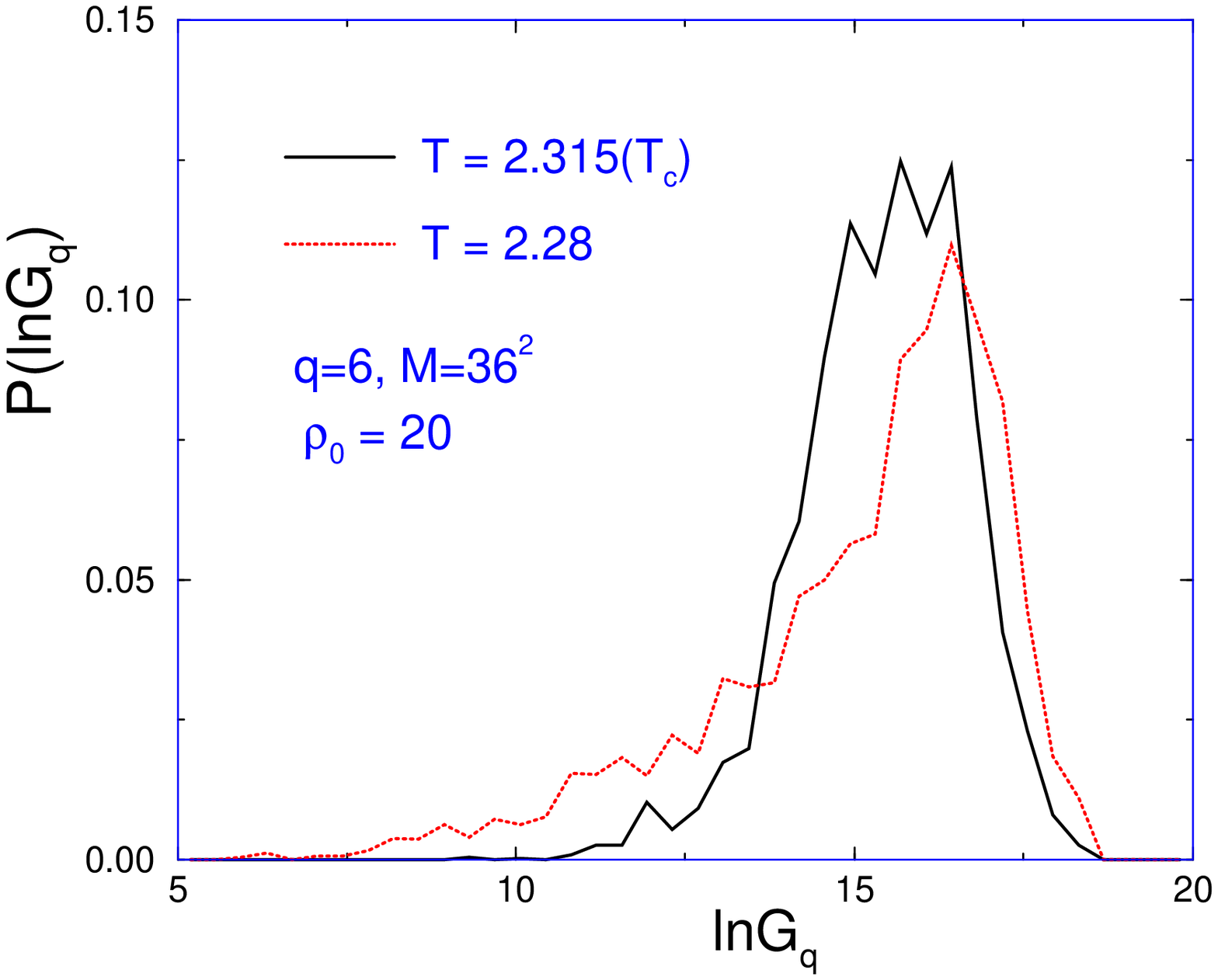}}
\vskip 1.5 cm
\centerline{\bf \large ~~~~~~~~~~~~~~Fig.3}
\vskip -0.5 cm
\end{figure}

\newpage
\begin{figure}[t]\epsfxsize=16.5cm \epsfysize=14cm
\centerline{\epsfbox{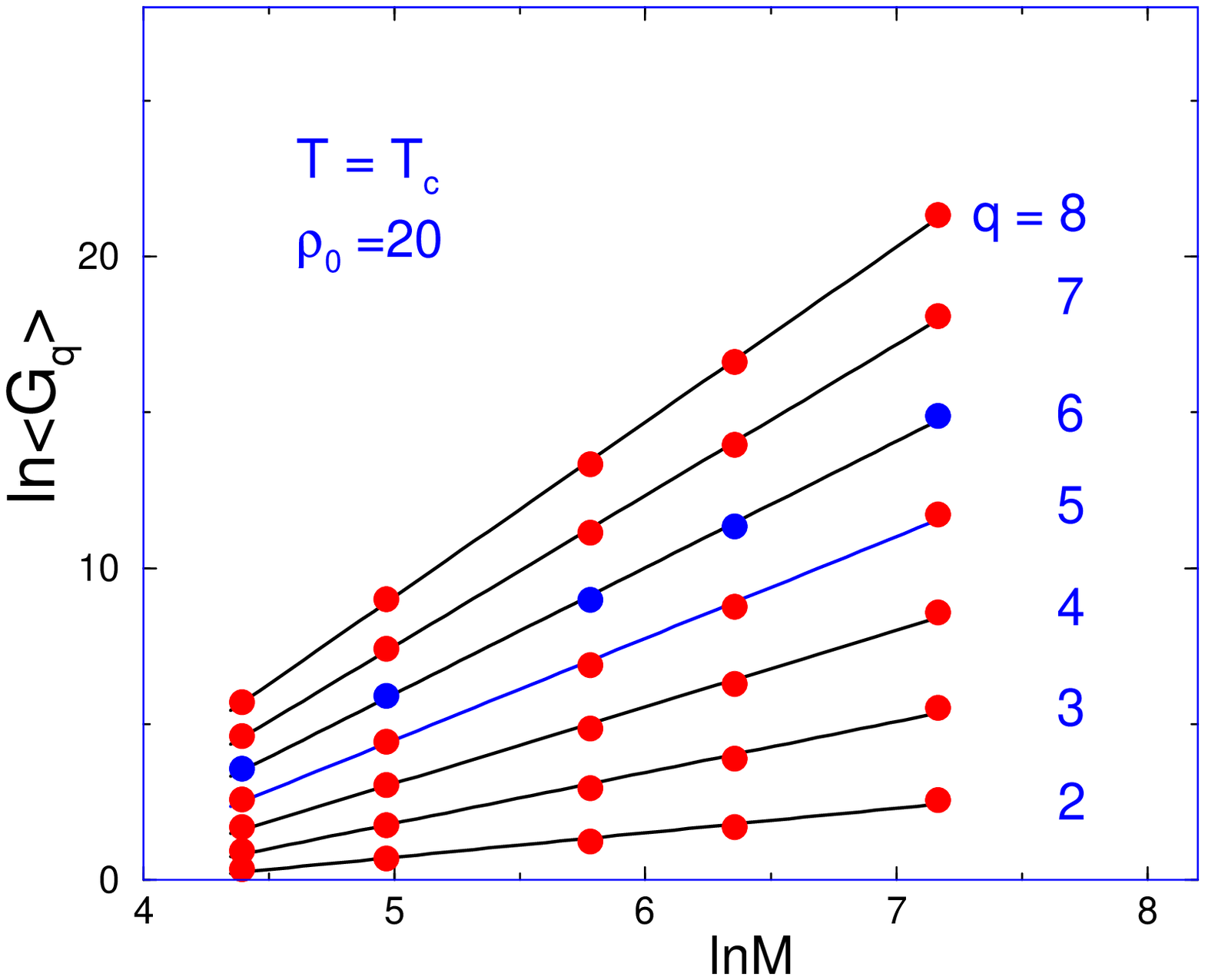}}
\vskip 1.5 cm
\centerline{\bf \large ~~~~~~~~~~~~~Fig.4}
\vskip -0.5cm
\end{figure}

\newpage
\begin{figure}[t]\epsfxsize=17cm \epsfysize=14cm
\centerline{\epsfbox{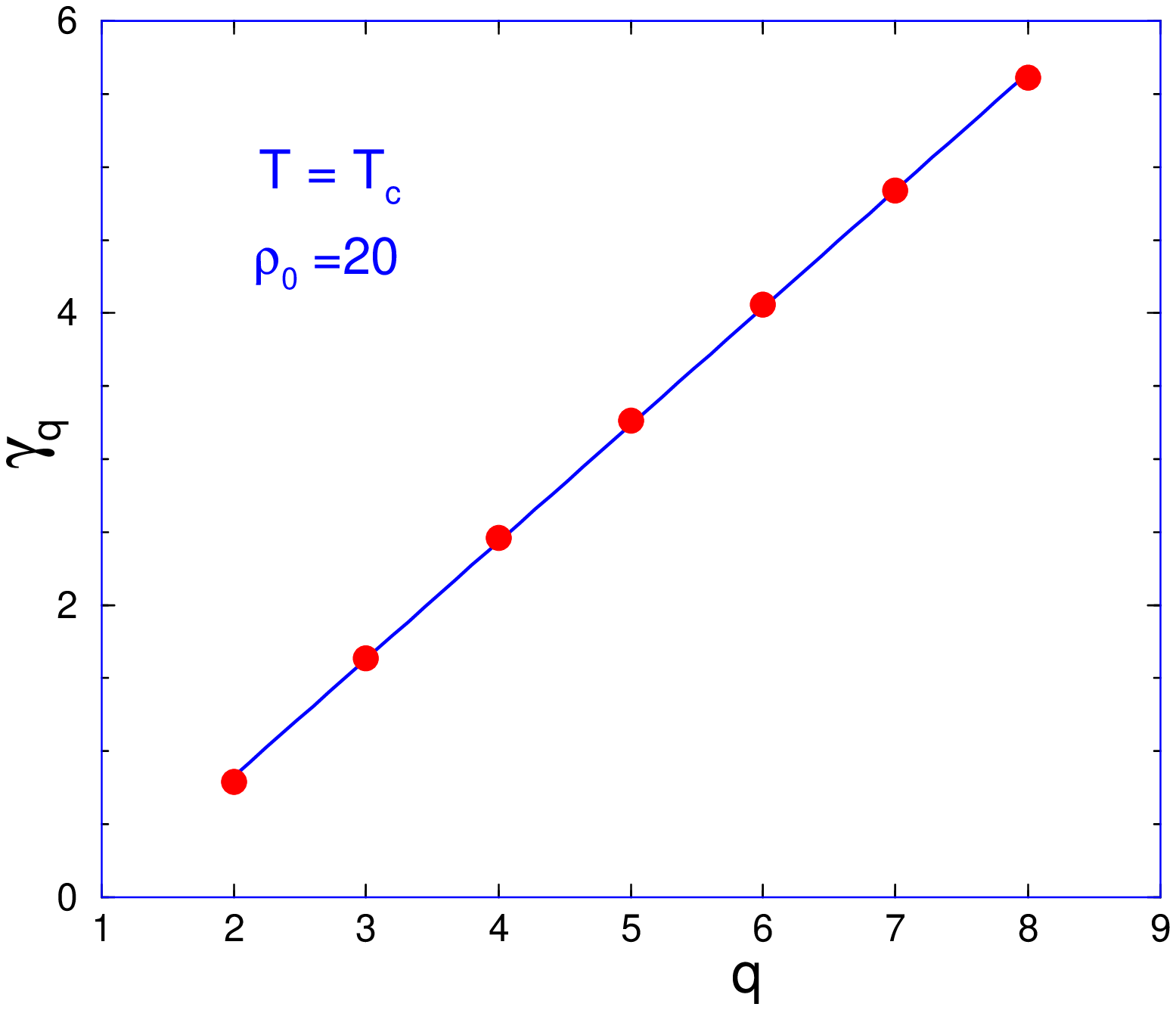}}
\vskip 1.5 cm
\centerline{\bf \large ~~~~~~~~~~~~~Fig.5}
\vskip -0.5cm
\end{figure}

\newpage
\begin{figure}[t]\epsfxsize=17cm \epsfysize=14cm
\centerline{\epsfbox{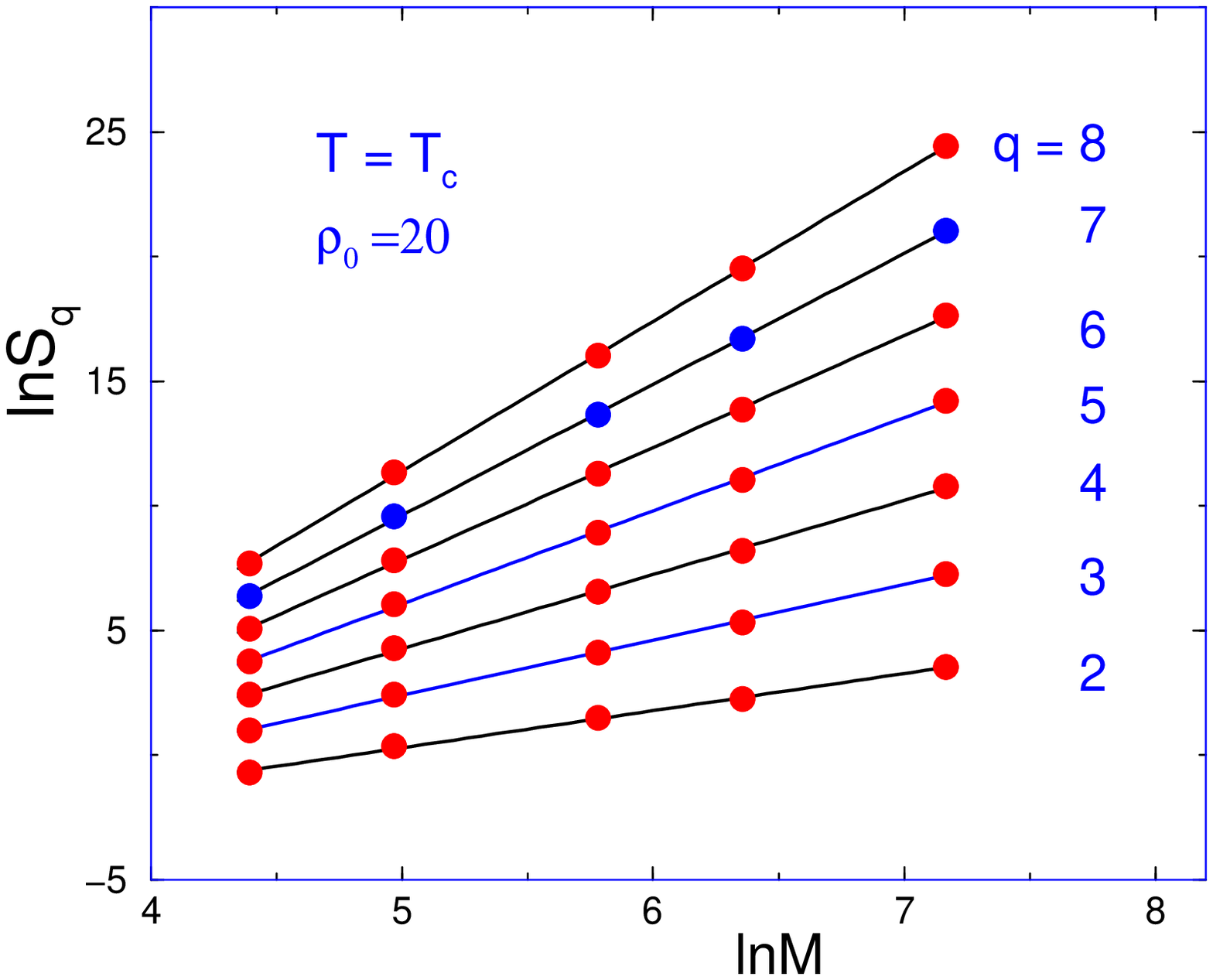}}
\vskip 1.5 cm
\centerline{\bf \large ~~~~~~~~~~~~~Fig.6}
\vskip -0.5cm
\end{figure}
\newpage
\begin{figure}[t]\epsfxsize=17cm \epsfysize=14cm
\centerline{\epsfbox{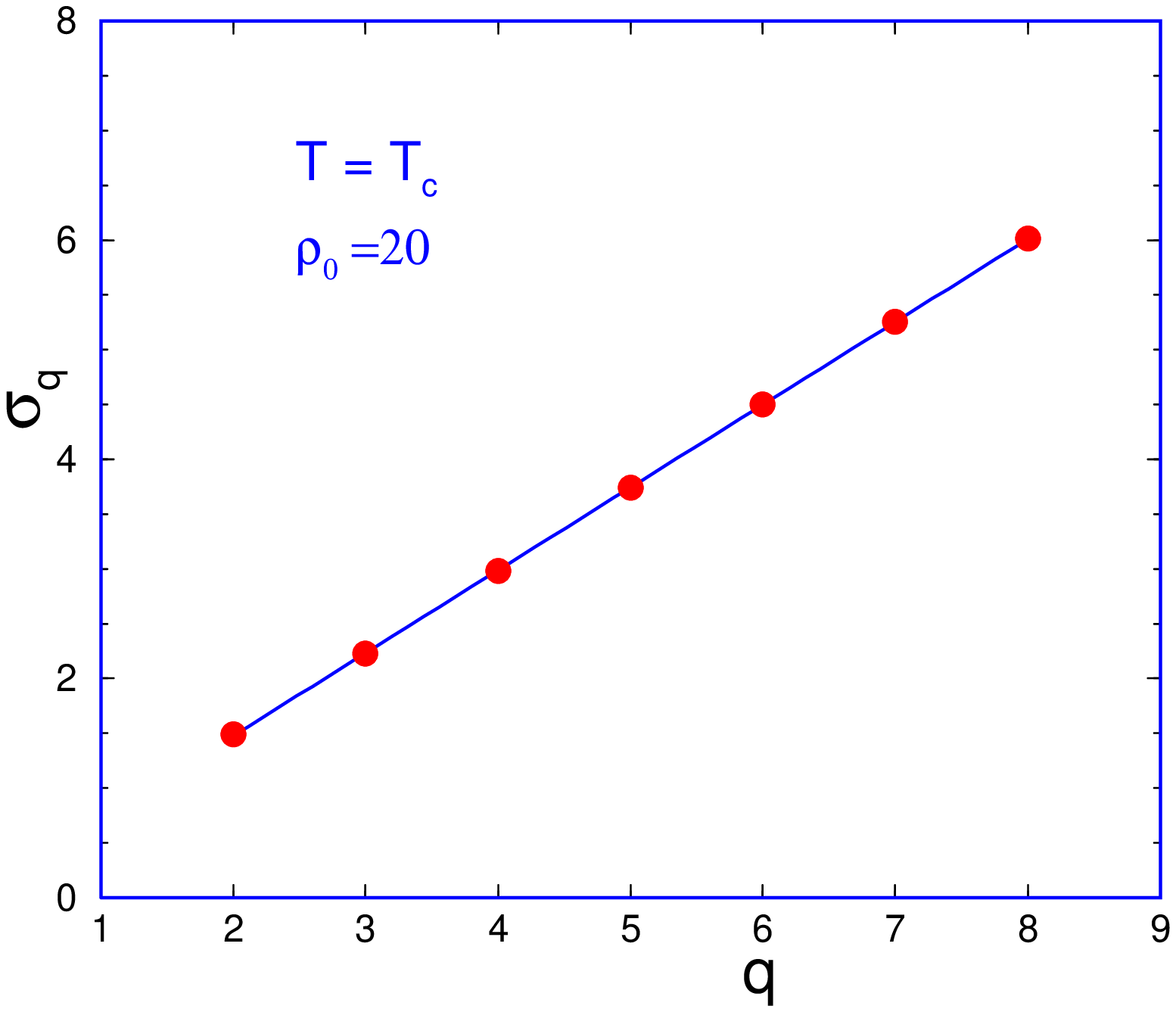}}
\vskip 1.5 cm
\centerline{\bf \large ~~~~~~~~~~~~~Fig.7}
\vskip -0.5cm
\end{figure}
\newpage
\begin{figure}[t]\epsfxsize=17cm \epsfysize=14cm
\centerline{\epsfbox{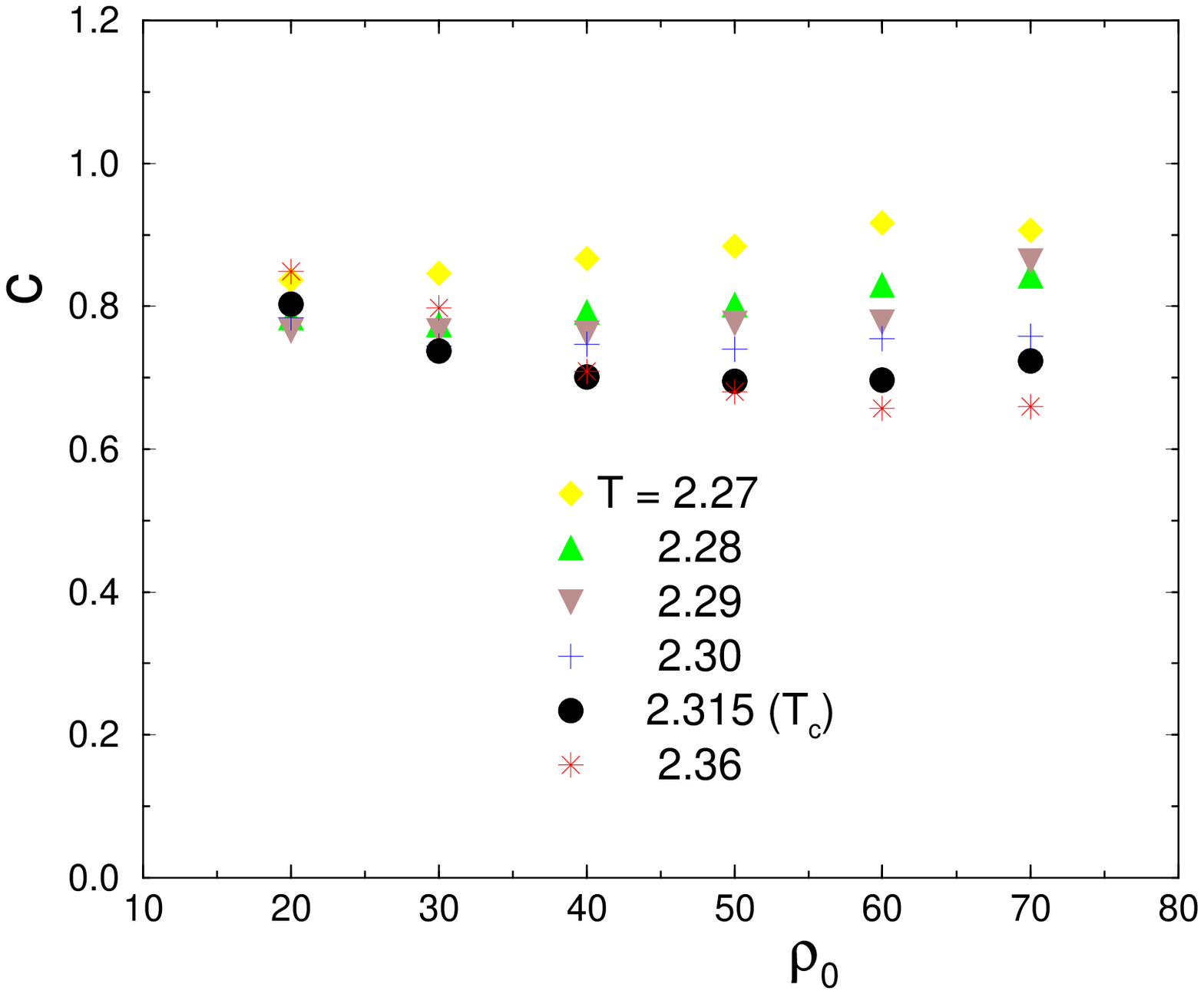}}
\vskip 1.5 cm
\centerline{\bf \large ~~~~~~~~~~~~~Fig.8}
\vskip -0.5cm
\end{figure}
\newpage
\begin{figure}[t]\epsfxsize=17cm \epsfysize=14cm
\centerline{\epsfbox{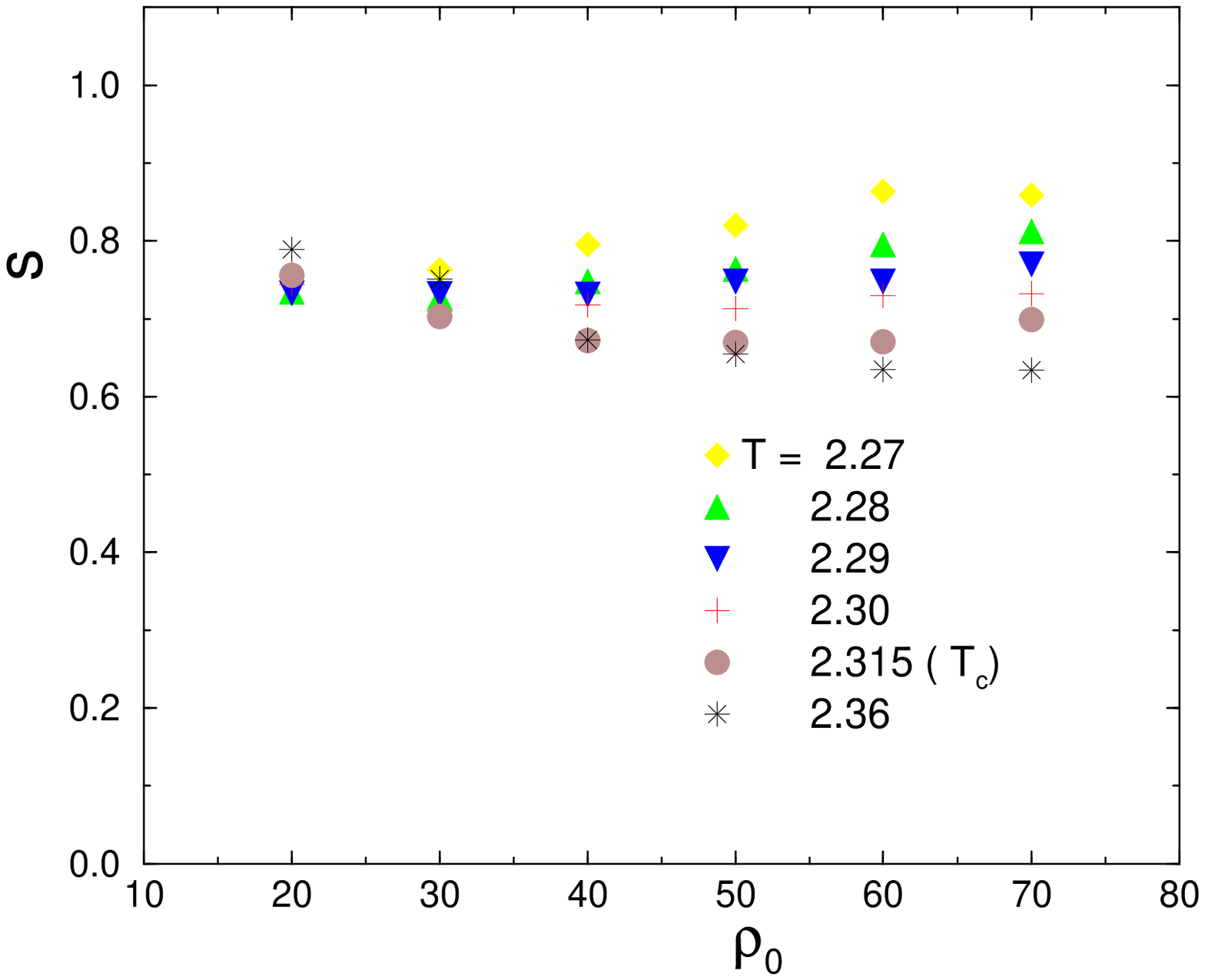}}
\vskip 1.5 cm
\centerline{\bf \large ~~~~~~~~~~~~~Fig.9}
\vskip -0.5cm
\end{figure}

\newpage
\begin{figure}[t]\epsfxsize=17cm \epsfysize=14cm
\centerline{\epsfbox{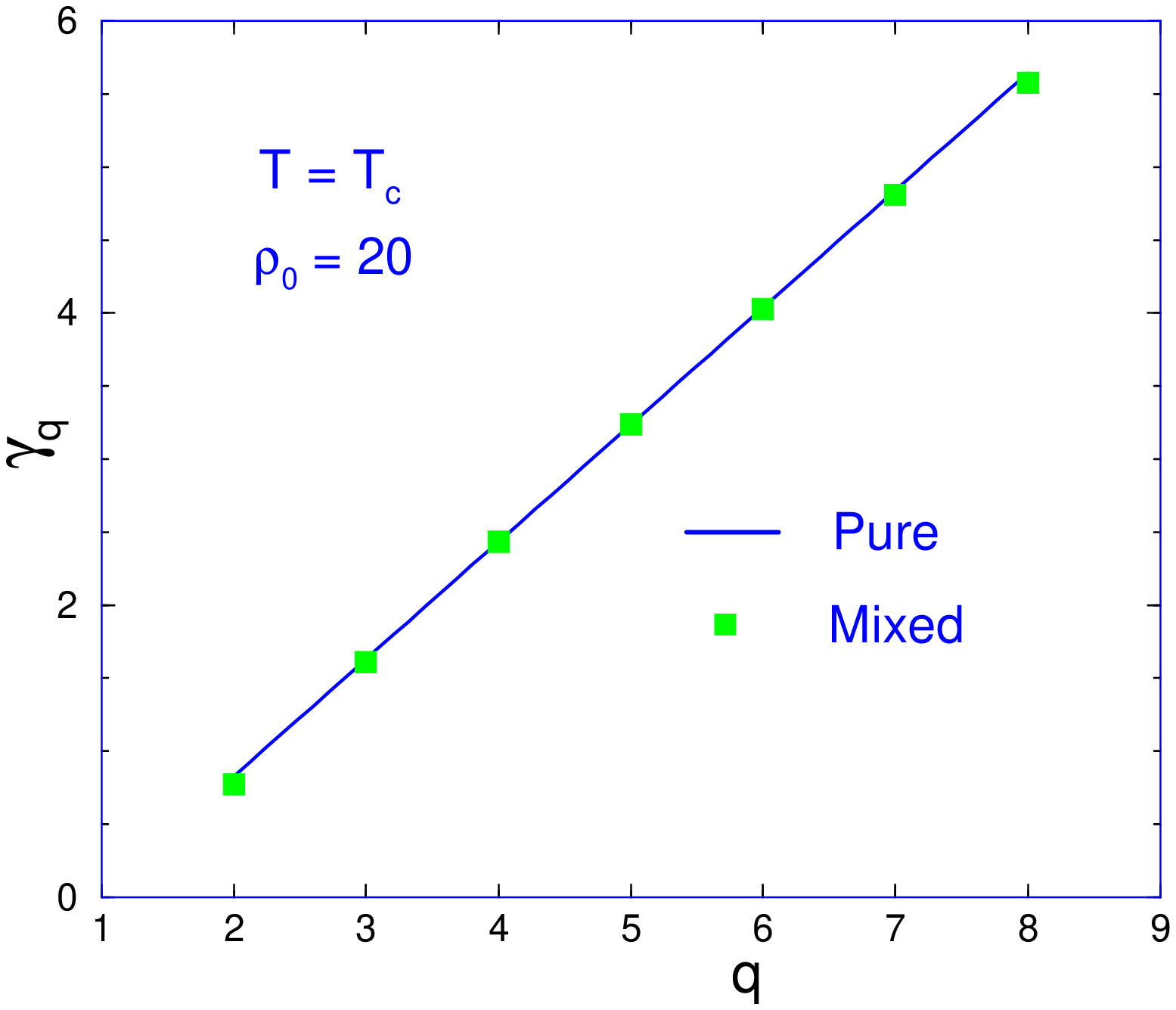}}
\vskip 1.5 cm
\centerline{\bf \large ~~~~~~~~~~~~~Fig.10}
\vskip -0.5cm
\end{figure}
\newpage
\begin{figure}[t]\epsfxsize=17cm \epsfysize=14cm
\centerline{\epsfbox{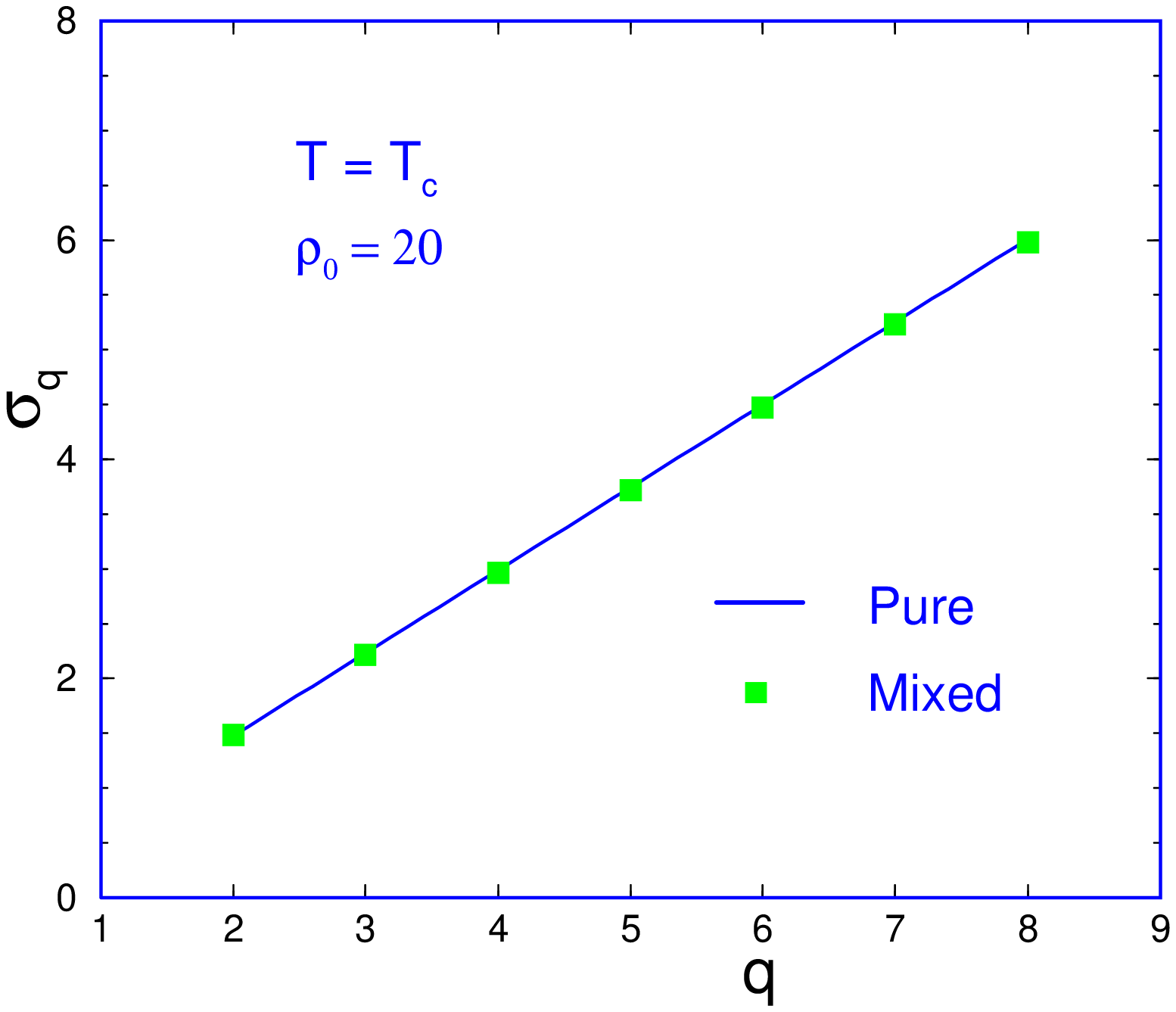}}
\vskip 1.5 cm
\centerline{\bf \large ~~~~~~~~~~~~~Fig.11}
\vskip -0.5cm
\end{figure}



\newpage
\begin{thebibliography}{000}

\bibitem{1}
R.\ C.\ Hwa and Y.\ F.\ Wu, Phys.\ Rev.\ C {\bf 60}, 054904
(1999).

\bibitem{2}J.\ J.\ Binney, M.\ J.\ Dowrick, A.\ J.\ Fisher, and M.\ E.\ J.\
Newman, {\it The Theory of Critical Phenomena} (Claredon Press, Oxford, 1992).

\bibitem{3}R.\ C.\ Hwa and Q.\ H.\ Zhang, hep-ph/9912275, Phys.\ Rev.\ D
(to be published).

\bibitem{4}R.\ Graham and H.\ Haken, Z.\ Physik {\bf 213}, 420 (1968); {\bf
237},
31 (1970);  V.\ DeGiorgio and M.\ Scully, Phys.\ Rev.\ A{\bf 2}, 1170
(1970); R.\ C.\ Hwa, Acta Physica Slovaca {\bf 49}, 201 (1999).

\bibitem{5}M.\ R.\ Young, Y.\ Qu, S.\ Singh and R.\ C.\ Hwa, Optics
Comm.\ {\bf 105}, 325 (1994).

\bibitem{6}R.\ C.\ Hwa and M.\ T.\ Nazirov, Phys.\ Rev.\ Lett.\ {\bf
69}, 741 (1992)

\bibitem{7}Z.\ Cao, Y.\ Gao and R.\ C.\ Hwa, Z.\ Phys.\ C {\bf 72}, 661
(1996).

\bibitem{8}See C.\ DeTar in {\it Quark-Gluon Plasma 2}, edited by R.\ C.\
Hwa (World Scientific, Singapore, 1995).

\bibitem{9}See K.\ Rajagopal, {\it loc.\ cit.}

\bibitem{10}R.\ C.\ Hwa, Phys.\ Rev.\ D {\bf 41}, 1456
(1990).

\bibitem{11}
U.\ Wolff, Phys.\ Rev.\ Lett.\ {\bf 62}, 361 (1989).

\bibitem{12}
Z.\ Cao and R.\ C.\ Hwa, Phys.\ Rev.\ Lett.\ {\bf 75},
1268 (1995); Phys.\ Rev.\ D {\bf 53}, 6608 (1996);  {\it ibid} {\bf 54},
6674 (1996); Phys.\ Rev.\ E {\bf 56}, 326
(1997).

\end{thebibliography}
\end{document}